\documentclass[%
reprint,
amsmath,
amssymb,
aps,
]{revtex4-2}
\usepackage{graphicx}
\usepackage{dcolumn}
\usepackage{bm}
\usepackage{amsmath}
\usepackage{hyperref}

\begin{document}

\preprint{APS/123-QED}

\title{Challenges in identifying simple pattern-forming mechanisms\\ in the development of settlements using demographic data}
\thanks{Third version}%

\author{Bartosz Prokop}
 \altaffiliation{Corresponding author: bartosz.prokop@kuleuven.be}
\affiliation{%
 Laboratory of Dynamics in Biological Systems,\\ Department of Cellular and Molecular Medicine, \\KU Leuven, Leuven, Belgium
}%

\author{Peter F. Pelz}
\affiliation{
 Chair of Fluid Systems\\
 TU Darmstadt, Darmstadt, Germany
}%

\author{Lendert Gelens}
\affiliation{%
 Laboratory of Dynamics in Biological Systems,\\ Department of Cellular and Molecular Medicine, \\KU Leuven, Leuven, Belgium
}%

\author{John Friesen}
\affiliation{
 Chair of Fluid Systems\\
 TU Darmstadt, Darmstadt, Germany
}%


\date{\today}

\begin{abstract}
\begin{description}
\item[Abstract]

The rapid increase of population and settlement structures in the Global South during recent decades motivates the development of suitable models to describe their formation and evolution. Such settlement formation has been previously suggested to be dynamically driven by simple pattern-forming mechanisms. Here, we explore the use of a data-driven white-box approach, called SINDy, to discover differential equation models directly from available spatiotemporal demographic data for three representative regions of the Global South. We show that the current resolution and observation time of the available data is insufficient to uncover relevant pattern-forming mechanisms in settlement development. Using synthetic data generated with a generic pattern-forming model, the Allen-Cahn equation, we characterize what the requirements are on spatial and temporal resolution, as well as observation time, to successfully identify possible model system equations. 
Overall, the study provides a theoretical framework for the analysis of large-scale geographical/ecological systems, and it motivates further improvements in optimization approaches and data collection.
\end{description}

\end{abstract}

\maketitle

\section{Introduction}
\label{sec:Introduction}
The fraction of the population living in urban or settlement structures has grown exponentially over the last several decades, especially in the Global South \cite{UN.2019, UN2020}.
This trend poses one of the main challenges in our world \cite{Retief2016} as the rising population in such structures is in need of vital infrastructure \cite{Adams2020} while simultaneously affecting (mostly negative) climatic developments  \cite{Nagendra2018,Thacker2019}.
Consequently, there is an urgent need to understand underlying processes of urbanization and anticipate the emergence of these structures.

\begin{figure}[b]
    \centering
    \includegraphics[width=8.5cm]{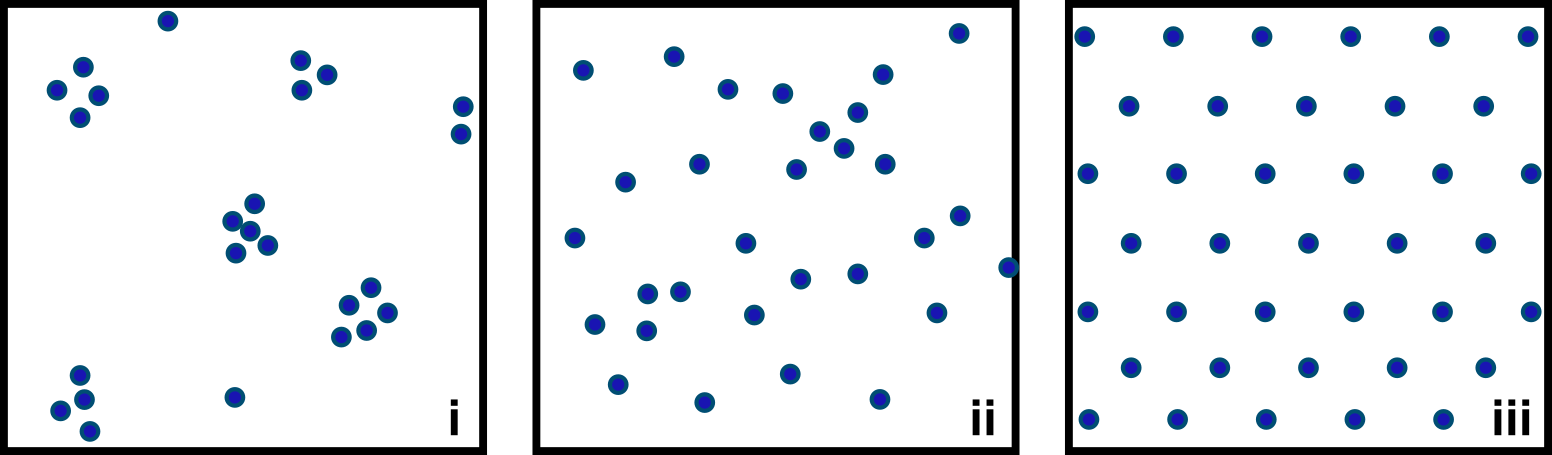}
    \caption{Possible settlement arrangements following the phenomenological interpretation as point processes, resulting in either clustered (i), random (ii) or regular distributions (iii).}
    \label{fig:distributions}
\end{figure}

Urbanization and development of settlement structures depends on several mechanisms based on repulsion and attraction \cite{Christaller1933,Hudson1969}.
Such interactions can lead to three major settlement distributions (see Fig.~\ref{fig:distributions}). Their existence has been confirmed in recent settlement pattern studies of different regions in the Global South, in which regular distributions are dominating \cite{Yang2016,AbouKorin2018, Friesen2018, Henn2020, Prokop2021}.

The existence of regularly-patterned distributions in settlements, and in other spatial systems, can be an indicator for the existence of instability-driven pattern-forming mechanisms \cite{Pringle2017}. A similar concept of linking spatial distributions with specific driving mechanisms has been successfully applied in the field of plant and animal ecology for decades \cite{Theraulaz2002, Grohmann2010,Tarnita2017}.
To understand the emergence of these patterns from underlying interactions, different modeling approaches have been developed. 
For example, urban development can be modeled by agent- or cellular-automata-based approaches \cite{Losiri2016, Batty2021} which include detailed interactions at the level of individuals.
Despite their accuracy in specific cases, such models have several drawbacks. 
They lack generalization, require large and detailed datasets to be fit on and can become computationally expensive. 
Another approach is to use reaction-diffusion models. Such systems consisting of differential equations are simpler and directly interpretable, yet they can also lead to highly complex patterns and dynamics \cite{May1976,Turing1990}. 
In the context of urban structures, Pelz \textit{et al.} \cite{Pelz2019} have developed a theoretical framework describing the formation of informal settlements (so-called \textit{slums}) in the Global South. Furthermore, this framework was extended to describe the morphogenesis of urban systems in the United States as a reaction-diffusion system in \cite{Friesen2019}.

Deriving such models is typically done by suggesting sets of possible models from experimental measurements of specific interactions paired with scientific intuition and determining which one provides the best fit to measured data. 
However, as data structures of urban system are complex, suggesting certain model structures can be difficult. 
One way to address this is by using recently developed data-driven model discovery approaches, such as the system identification approach called "Sparse Identification of Nonlinear Dynamics" (SINDy) \cite{Brunton2016}.
SINDy has  recently gained attention in many fields, such as engineering \cite{Reinbold2021}, physics \cite{Ermolaev2022}, chemistry \cite{Hoffmann2019} and biology \cite{Mangan2016}. 
The method has shown success in identifying interpretable models in the form of ordinary or partial (through the extension PDE-FIND \cite{Rudy2017}) differential equations from synthetic data of e.g.\ pattern-forming mechanisms in the form of reaction-diffusion equations \cite{Rudy2017,Schaeffer2017}. 
However, the literature on model discovery from real temporal or spatiotemporal data with SINDy is scarce (e.g.\ on generic benchmark problems in \cite{Fasel2022} or \cite{Hirsh2022}), as the method struggles with commonly encountered high-noise and low-data situations. 

In this work, we will theoretically and practically investigate if pattern-forming processes can be responsible for settlement development in the Global South and attempt to identify such mechanisms using SINDy directly from existing spatiotemporal data of population patterns.
In section \ref{sec:settlement_pattern}, we motivate the potential relevance of pattern-forming equations from data analysis of satellite images.
In section \ref{sec:sindy_settlements}, we apply the SINDy algorithm to satellite images and try to identify mathematical equations in the form of one-component pattern-forming mechanisms. 
As model identification from the real data sets turns out to be challenging, we identify and study two main challenges, namely data availability and quality and determine their influence on model discovery from spatiotemporal data in section \ref{sec:sindy_AC}. 
Lastly, we discuss our findings and elaborate on these challenges and what is required to overcome them in section \ref{sec:discussion}.

\section{Settlement development as a pattern-forming process}
\label{sec:settlement_pattern}

We start our study by selecting three representative regions of emerging countries ~--~ the Punjab region in India, the Nile delta in Egypt and the Kano State region in Nigeria (see Table~\ref{tab:regions}).
The regions are chosen as they lie in countries which can be considered representative of the Global South:
all countries have had steady population as well as steady economic growth over the last 20 years \cite{UN.2019}.
Despite and because of the great cultural differences, all three societies are in transition from agricultural countries to industrialized nations. 

\begin{table}[b]
\caption{\label{tab:regions} Investigated regions with their geopolitical location and attributes.}
\begin{ruledtabular}
\begin{tabular}{lll}
 Region  & Attributes \\ \colrule
 Punjab, India& northwest India, border region with\\
 &Pakistan, agricultural region called\\
 &Granary of India \\
 Nile delta, Egypt & north Egypt, densely populated, fast\\
 &growing agricultural region \\
 Kano State, Nigeria & north Nigeria, border region to Niger,\\
 &agricultural region in one of the fastest\\
 &growing economies
\end{tabular}
\end{ruledtabular}
\end{table}

\begin{figure*}
	\includegraphics[width=0.8\linewidth]{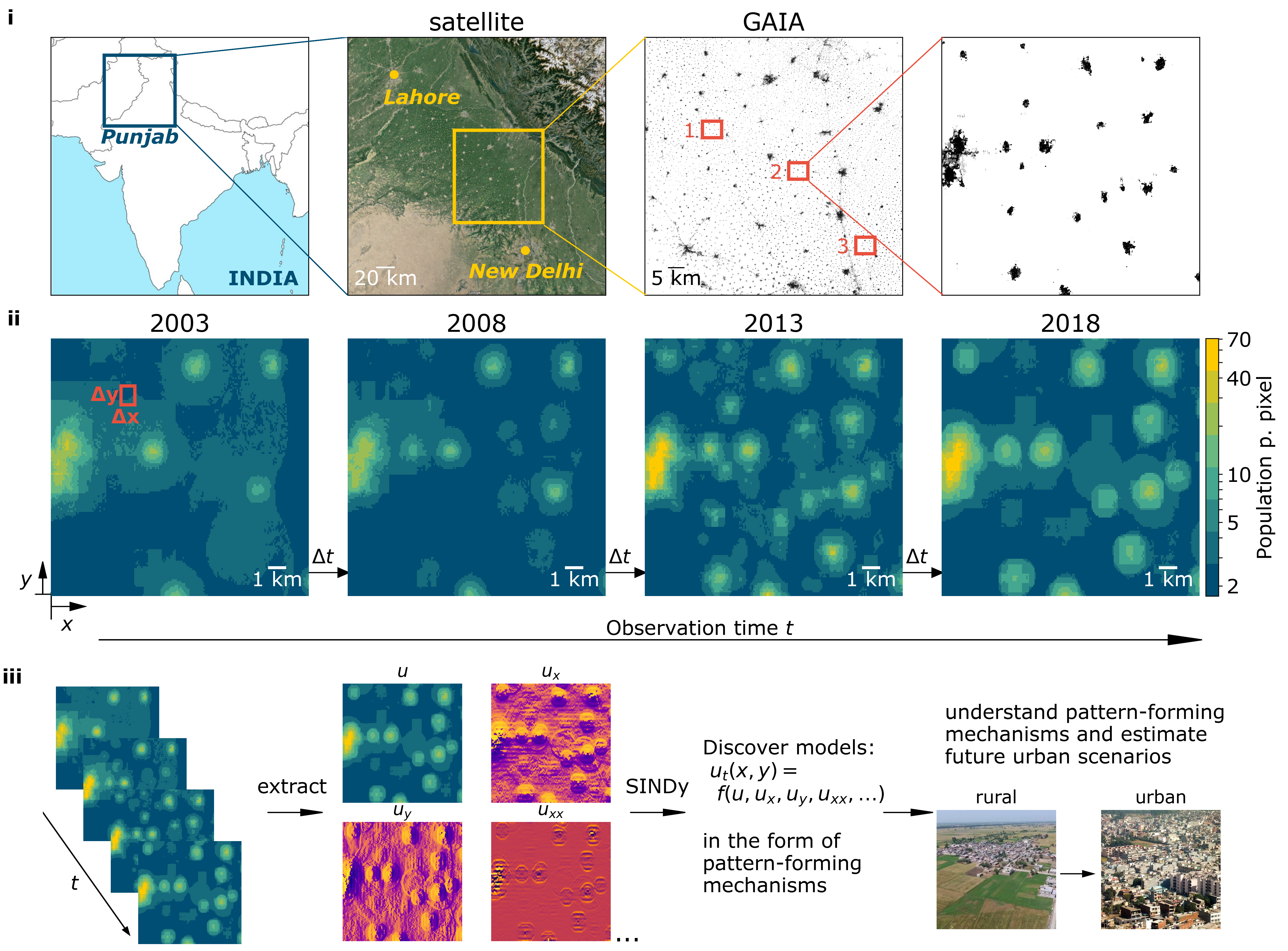}
	\caption{Overview of the workflow of this study i) Selection of ROIs, starting from the identification of suitable regions of the Global South, here Punjab, India, selection of an excerpt of the region in the data set GAIA and determination of ROIs from regular settlement patterns with the ANN (see Appendix~\ref{app:MW}). ii) Data of population density patterns (on logarithmic scale) from the data set \textit{WorldPop} from ROI 2 of Punjab, India, for four time points with a spatial resolution of $\Delta x,\Delta y\approx100$m (enlarged in figure) at the equator and temporal resolution of $\Delta t=1 \text{ year}^{-1}$ iii) Work flow to investigate possible pattern-forming mechanisms in the development of settlement structures. Starting from the time series of population density patterns, we extract spatiotemporal features and apply SINDy to discover models in the form of one-component partial differential equations. This can help us understand the role of such mechanisms in settlement structures and estimate future urban scenarios. }
	\label{fig:data_selection}
\end{figure*}

Furthermore, the spatial distribution of settlements has been studied in these regions and their regularity characterized on different spatial scales \cite{AbouKorin2018,Henn2020,Prokop2021}. This allows us to select subregions (here called regions of interest (ROI)) which show a regular distribution. This is illustrated for the Punjab region in India in Fig.~\ref{fig:data_selection}i). For a detailed explanation of how we used satellite data, in combination with the \textit{Global Artificial Impervious Area} (GAIA, temporal resolution $\Delta t = 1 \text{ year}^{-1}$, spatial resolution $\Delta x = \Delta y \approx 30$ m) \cite{Gong2020,Liu2020} data set, to select the ROIs, we refer to Appendix \ref{app:MW}. For each ROI, we then track the settlement evolution over a period of about 15 years using the data set \textit{WorldPop} \cite{Worldpop} depicting spatial population distributions, see Fig.~\ref{fig:data_selection}ii. The \textit{WorldPop} data set is used later for the model identification process as described in Fig.~\ref{fig:data_selection}iii 
\footnote{The computation of the ANN, the resulting selection of the ROIs, and the later computation of the feature lengths was done with GAIA because it provides discrete spatial data (1 - built-up, 0 - no built-up area). As such, and in contrast to \textit{WorldPop}, which distributes population quantities across administrative regions, leading to non-existent values of 0, the identification of settlement locations and sizes is straightforward. However, for model identification \textit{WorldPop} is used, since most pattern forming mechanisms describe concentration distributions of different (chemical or other) species and do not create discrete patterns \cite{Murray2003,Cross1993}, similar to \textit{WorldPop}.}.

Previous work has shown that the observed emergence of rural settlement structures could be caused by simple reaction-diffusion pattern-forming mechanisms \cite{Pelz2019}. Following the example of Pelz \textit{et al.}, we divide the system into a population living in the rural settlements, and a supply potential of agricultural land which is complementary to the population. In other words, areas that are in agricultural use are uninhabited, and vice-versa. We also assume that there is a limit to agricultural exploitability and urban densification. The population density at a spatial point $\bm{x}=(x,y)$ and time $t$ is given by $u(\bm{x},t)$ and the corresponding supply potential by $v(\bm{x},t)$.

The change of the concentrations of $u$ and $v$ is defined by three complementary global contributions: (i) birth or death of the population or the cultivation or sealing of agricultural space within a domain of size $A$, (ii) migration to and from other cities outside of $A$ leading to a de- or increase in supply potential and (iii) migration to areas with higher supply potential over the boundary $C$ of $A$ where a settlement exists or is created. These mechanisms lead to the formulation of a typical reaction-diffusion model (for more detail, see Appendix~\ref{app:rd_equations} and \cite{Pelz2019}): 
\begin{align}
	\begin{split}\label{eq:RD_general}
		&u_t =  \nabla^2 u + R f(u,v)\\
		&v_t = D \nabla^2 v + R g(u,v),\\ 
		&\text{with} \quad (\bm{n}\cdot \nabla) \begin{pmatrix} u	\\	v \end{pmatrix}=0 \text{ on } C,
	\end{split}
\end{align}
with $D,R$ being the diffusion and reaction coefficients alongside the respective reaction terms $f,g$, which are being evaluated under no-flux boundary conditions.

The linear stability analysis of the homogeneous system (no diffusion) provides the Jacobian matrix $\bm{J}$.
It is well known that specific sign-combinations of components of the Jacobian allow for a initial homogeneous distribution of population and supply potential to be stable ~--~ either through substrate inhibition or as an activator-inhibitor system \cite{Murray2003}.
In the case of settlement structures, the only physically sensible choice is substrate inhibition, as the resulting concentration patterns are out of phase \cite{Pelz2019,Murray2003}:

\begin{figure*}
	\centering
	\includegraphics[width=0.99\linewidth]{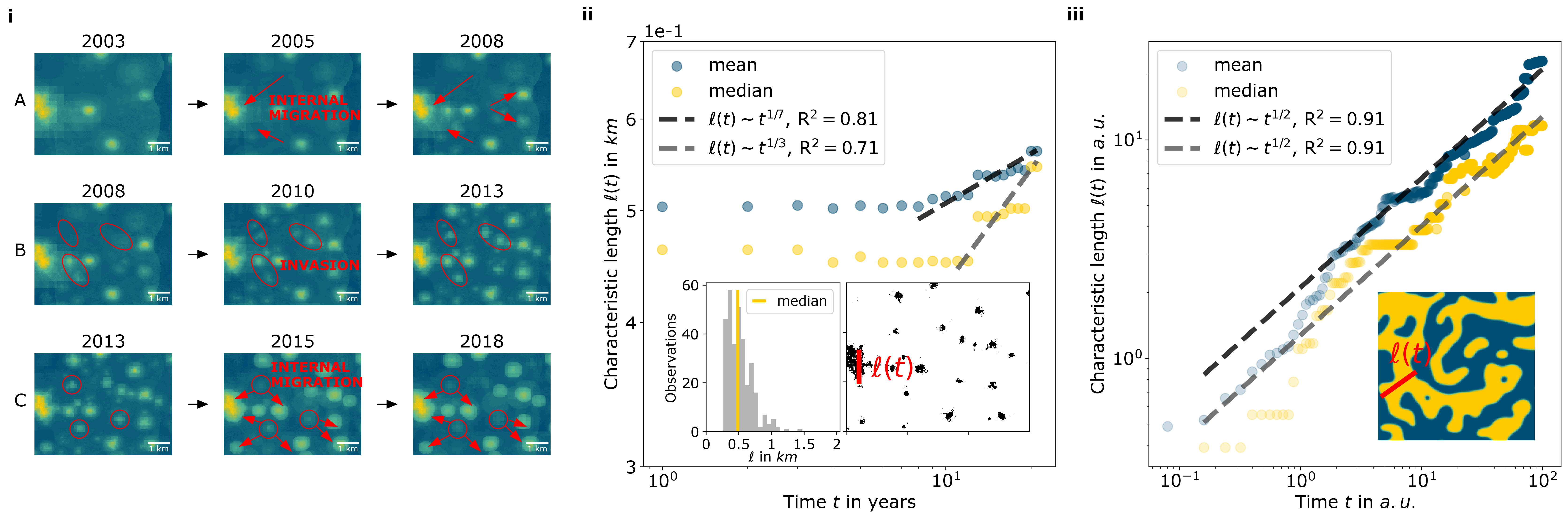}
	\caption{(i) Example of multiple effects found in spatiotemporal data sets of population densities indicating the existence of suggested reaction-diffusion equations (population density data from data set \textit{WorldPop} \cite{Worldpop}, ROI 2, Punjab, India). The observable behaviors are internal migration into existing settlements (A), invasion of not occupied agricultural space (B) or local migration triggered by competition over available agricultural space (C). (ii) Calculated characteristic length of settlement patterns resulting from the mean and the median heavy sided feature size distribution (shown in inlet, ROI 2, Punjab, India in 2018). The characteristic lengths obey the power law when growing, with $\ell(t) \sim t^{1/7}$ and $\ell(t) \sim t^{1/3}$ respectively (with R$^2$ scores). iii) Calculated characteristic length of the pattern (example of a feature size shown in inlet) resulting from the Allen-Cahn equation. Here, both the mean and median characteristic length obey the power law with $\ell(t) \sim t^{1/2}$ when growing (with R$^2$ scores).}
	\label{fig:dynamics}
\end{figure*}

\begin{equation}
	\bm{J} = \begin{pmatrix}  f_u\big|_0 & f_v\big|_0\\ g_u\big|_0 & g_v\big|_0  \end{pmatrix}= \begin{pmatrix}  + & + \\ - & -  \end{pmatrix}.
\end{equation}

\begin{description}
	\item[$f_u\big|_0 > 0,$ \textit{people attract other people}] The population $u$ increases due to self-reproduction of $u$ in an environment of sufficient sustenance. 
	\item [$f_v\big|_0 > 0,$ \textit{supply attracts people}] The amount of available agricultural area $v$ attracts people, increasing the population $u$ of the settlement.
	\item [$g_u\big|_0 < 0,$ \textit{people inhibit supply}] The higher the population $u$, the less agricultural area $v$ is available, especially due to the limitation by the agricultural needs of surrounding settlements. This is the case until the maximum amount of agricultural area is used and does not suffice to supply the population of a settlement, leading to a decrease of population $u$.
	\item [$g_v\big|_0 < 0,$ \textit{supply inhibits additional supply}] Due to limited resources, the supply production decreases, when agricultural efficiency plateaus.
\end{description}

In the presence of diffusion, the linear stability shows that an equally dispersed population and supply potential densities can destabilize to form spatial patterns when the following condition is met (see Appendix~\ref{app:LinStabRD}): 
\begin{equation}
	f_u\big|_0 > - \frac{g_v\big|_0}{D}.
\end{equation}
This is the case when the attraction of people to $A$ dominates the inhibition of supply potential due to the emergence of settlements. If the domain size $A$ and diffusion coefficient $D$ of the system are sufficiently large, different settlements can emerge that constantly compete against each other over the supply potential, eventually leading to a regular distribution of settlements.

Assuming that the total population density $u$ and supply potential $v$ are conserved, $u + v = c_{max}$, the system reduces to a simpler one-component equation \cite{Cahn1958, Cross1993}:
\begin{align}
	\begin{split}
		u_t(x,y,t) &= \check{D} \nabla^2 u(x,y,t) + R \check{f}(u(x,y,t)) \label{eq:RD_eq}, \\ 
		\text{with} \quad \check{D} &= \frac{D+1}{2}, \\
		\text{and  } \quad \check{f} &= \frac{1}{2}\left[f(u(x,y,t))+g(u(x,y,t))\right].
	\end{split}
\end{align}

Based on this proposed reaction-diffusion description consistent with pattern formation \cite{Pelz2019}, we first set out to see whether the satellite data (Fig.~\ref{fig:data_selection}ii) contained any clear signatures of such developing rural settlement patterns. Visual inspection of the \textit{WorldPop} data set indicates the possible presence of three key processes: local migration through attraction of bigger settlements (e.g. between 2003 and 2008, Fig.~\ref{fig:dynamics}i/A), invasion or occupation of available agricultural space (e.g. between 2008 and 2013, Fig.~\ref{fig:dynamics}i/B), and local migration induced by competition of settlements over available space leading to changes in settlement patterns (e.g. between 2013 and 2018, Fig.~\ref{fig:dynamics}i/C). 

We then analyzed the characteristic lengths $\ell(t)$ of settlements (see Fig.~\ref{fig:dynamics}ii) using the GAIA data set \cite{Gong2020,Liu2020}. 
We chose the characteristic length to be the size of the respective features, evaluating the distribution feature sizes and determining the predominant size following \cite{Puri2004} and \cite{Konig2021} (see inset in Fig.~\ref{fig:dynamics}ii, for more details see Appendix~\ref{app:cl_calc}). 
This analysis shows that when the characteristic size of settlements is growing, this growth is well described by a power law $ct^{\beta}$, with an exponent of $\beta=1/7$ for the mean and $\beta=1/3$ median characteristic lengths. 

Interestingly, from literature we know that the time evolution of characteristic lengths of patterns driven by coarsening mechanisms also follow such a power law. For example, for the Allen-Cahn equation, it was theoretically shown that the change of $\ell(t)$ is described by the power law with $\beta=1/2$ \cite{Puri2004,Christiansen2020,Konig2021}, while for the Cahn-Hilliard equation \cite{Cahn1958}, $\beta= 1/3$ \cite{Puri2004,Konig2021}. Indeed, we calculated the characteristic length for a simulated pattern for the Allen-Cahn equation of the form,
\begin{align}
	\label{eq:AC_org}
	\begin{split}
	    u_t(x,y,t)= \: &\alpha \nabla^2 u + \beta u + \gamma u^2 - u^3,
	\end{split}
\end{align}
using the same method as for the settlement patterns, which revealed that the mean and median characteristic length $\ell(t)$ follow the power law with $\beta=1/2$ (see Fig.~\ref{fig:dynamics}iii).
Even though the exponent $\beta$ of the power law for settlement patterns is not the same as for the Allen-Cahn or Cahn-Hilliard equation, it indicates that settlement patterns could be a product of a simple, coarsening pattern-forming mechanism in the proposed form of Eq.~(\ref{eq:RD_eq}).

\section{Model identification from settlement data}
\label{sec:sindy_settlements}

\subsection{Model identification using SINDy}

The SINDy method to identify differential equation models from spatiotemporal datasets has been increasingly applied in many fields, i.e. in fluid mechanics \cite{Brunton2016}. However, it has, to our knowledge, never been used for studying large scale, geo-sociological questions. Therefore, we asked ourselves whether we could use this method to discover a partial differential equation (PDE) that provides a good description of the measured time evolution of rural settlements in the Global South. If successful, such a PDE would also provide new insights in potential pattern-forming mechanisms in settlement development. 

The main idea behind the SINDy method is the assumption that dynamic systems can be described through either ordinary, or in our case partial differential equations (using PDE-FIND) \cite{Rudy2017}, with sparse structure in the following form:
\begin{equation}
    u_t = N (u(\bm{x},t), u_x, u_y, ..., \bm{x}, \bm{\xi} )
\end{equation}
The temporal change of $u$, $u_t$, is a function of the variable $u$ itself, its spatial derivatives and a set of coefficients $\bm{\xi}$. 
Differential equations of this form can be linearly combined: 
\begin{equation}
    u_t=\xi_1 + \xi_2 u + \xi_3 u^2 + \xi_4 u_x + \xi_5 u_{xx} + ...
\end{equation}
This equation can be rewritten as a row vector containing all combinations and derivatives of the quantity, called the term library and a coefficient vector $\bm{\xi}$ containing all coefficients:
\begin{equation}
    u_t = \begin{pmatrix} 1 & u & u^2 & u_x & u_{xx} & ... \end{pmatrix} \cdot \bm{\xi}.
\end{equation}
The values of each term in the library can be calculated from a single shot at a given point in time. 
If this system is extended to all available time points, a linear system of
equations with the unknown parameter vector $\bm{\xi}$ and the term library matrix $\bm{\Theta}$ is formed:
\begin{align}
\begin{split}
\label{eq:sindy_problem}
    \begin{pmatrix}  \\ \\ \bm{u}_t \\  \\\\\end{pmatrix} &= \begin{pmatrix}  &  &  &  &  & \\
     &  &  &  &  & \\1 & u & u^2 & u_x & u_{xx} & ... \\  &  &  &  &  &\\
     &  &  &  &  & \end{pmatrix} \cdot \begin{pmatrix}  \\ \bm{\xi} \\  \\  \end{pmatrix} =\bm{\Theta}\cdot\bm{\xi}
\end{split}
\end{align}

We assume that the settlement evolution could be captured by a PDE similar to Eqs.~(\ref{eq:RD_eq}) and (\ref{eq:AC_org}). Therefore, we use the term library given in Tab.~\ref{tab:TermLibs}, which includes derivatives up to the fourth order.

\begin{table}[b]
	\caption{\label{tab:TermLibs}%
		Terms included in the library for the one-component, two-dimensional equation sorted by combinations of $u$ and its derivatives.}
	\begin{ruledtabular}
		\begin{tabular}{p{1.7cm}p{6.3cm}}
			&Terms\\
			\colrule
			Combinations & $1$, $u$, $u^2$,$u^3$\\
			Derivatives & $u_{x}$,$u_{y}$, $u_{xx}$, $u_{yy}$, $u_{xy}$, $u_{xxx}$, $u_{yyy}$, $u_{xxy}$, $u_{yyx}$,\\
			&$u_{xxxx}$, $u_{yyyy}$, $u_{xxyy}$, $u_{xxxy}$, $u_{yyyx}$
		\end{tabular}
	\end{ruledtabular}
\end{table}

This system poses an over-determined optimization problem for values of $\bm{\xi}$ and can be solved using regression algorithms (for more detailed information on regression algorithms for SINDy, see \cite{Champion2020}).
In contrast to the original work in which the method PDE-FIND was introduced \cite{Rudy2017}, we apply a sparsity-promoting algorithm with the SR3 algorithm developed by \cite{Zheng2019}. This method includes the additional variable $\bm{w}$, which is forced to be close to the coefficient parameter and therefore relaxes the optimization problem. 

\begin{align}
	\begin{split}
	\min_{\bm{\xi},\bm{w}} &\frac{1}{2} \|\bm{u}_t- \bm{\Theta}\bm{\xi}\|^2_2 + \lambda \| \bm{w} \|_1 + \frac{\alpha}{2}\|\bm{w}- \bm{\xi}\|^2_2 \\ 
	&\text{with} \quad \lambda = \frac{l^2}{2\alpha}
	\end{split}
\end{align}

Here, two hyper-parameters of the optimization have to be set: the threshold $l$ and the parameter of the optimization $\alpha$, which provides the penalizing parameter $\lambda$ of the regularization. 

After model identification, we analyze the discovered models with the Akaike Information Criterion (AIC).
The AIC is a measure of parsimony \cite{Akaike1973}. 
It compares the goodness of fit of a given model to other proposed models and weighs it with the model's complexity aiming on maximizing the information provided by the simplest-as-possible model. 
For our analysis, we apply the corrected formulation for finite sample sizes of the AIC ($\text{AIC}_c$) proposed by \cite{Mangan2017}, resulting from \cite{Burnham2004}: 

\begin{equation}
    \text{AIC}_c= \text{AIC}+ \frac{2(k+1)(k+2)}{m-k-2}.
\end{equation}

The AIC is described by the likelihood function of average error over time and space $\epsilon$ as follows:

\begin{align}
\begin{split}
    \text{AIC} &= m \ln{(\epsilon/m)} + 2k \\
    \text{with } \quad \epsilon &=\frac{\left|\sum_{i=1}^{m_t} y_i - N (x_i, \bm{\xi} )\right|}{m} \\
    \text{and} \quad m &=m_s n_\text{ROI}
\end{split}
\end{align}

\begin{figure*}
	\centering
	\includegraphics[width=0.8\linewidth]{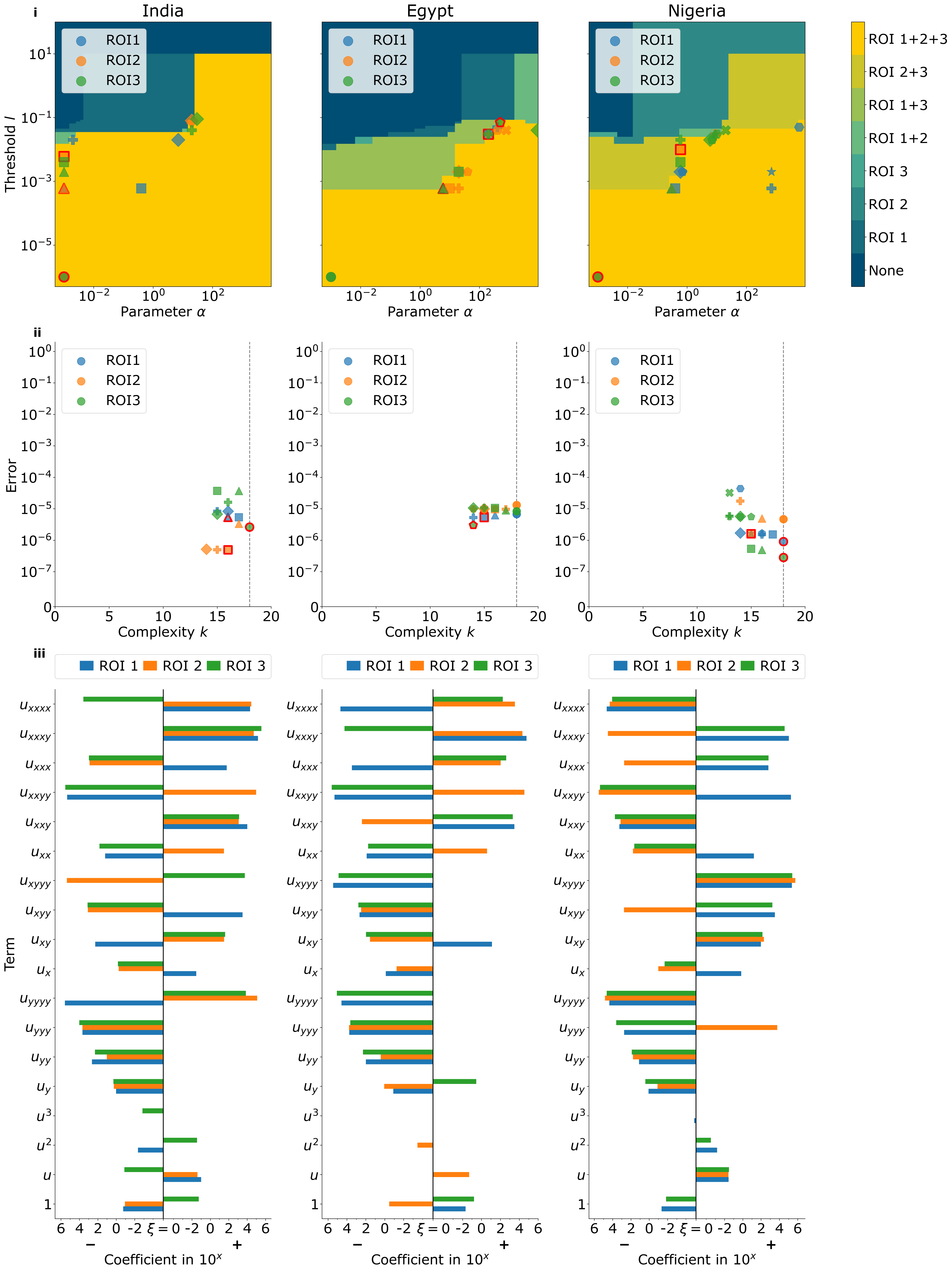}
	\caption{Parameter sweep of all ROIs and regions for sets of threshold $l$ and optimization parameter $\alpha$. (i) The analysis of the ROIs shows different combinations of parameters where the AIC falls beneath 2 for the respective ROI. As we search for a regionally valid solution, we select only unique solutions at combinations of $l,\alpha$ where the $ \text{AIC}<2 $ for all ROIs overlap. The selected unique equations and their respective optimization parameter sets where they were found first are shown. Red bordered markers depict the best identified models of each ROI. We see that the AIC provides the best equations with the lowest error and complexity (except for ROI 2, from which we show later through analysis of contributions that only the production terms are significant) (ii) The respective error and complexity of the found equations are depicted. The red bordered markers show the best model for each ROI in each region following the AIC. (iii) The identified coefficients $\bm{\xi}$ (as in Eq.~(\ref{eq:sindy_problem})) are shown for each equation. All identified models contain most of the derivatives (diffusive terms) with large coefficient values. If found the coefficient values of the reaction terms are multiple magnitudes smaller than of the diffusive terms.}
	\label{fig:results}
\end{figure*}

The AIC depends on the number of observations $m$ (size of region $m_s=XY$ and amount of included ROIs $n_{\text{ROI}}=3$, where we interpret an observation as the time series at every spatial point) and the number of terms ($k$) describing the complexity of an identified model.

Using this approach, we determine the most parsimonious model among all potential models and study its properties. 
In order to compare the identified models, we further normalize the AIC by the minimal value of the respective analysis $\text{AIC}_{min}$.
Here, following \cite{Mangan2017} and \cite{Burnham2004}, a model that has an $\text{AIC}_c-\text{AIC}_{min}<2$ has strong support for being the correct underlying system, while the ones with $\text{AIC}_c-\text{AIC}_{min}<8$ have weak support. Hereafter, we always refer to the corrected $\text{AIC}_c$ when the AIC is mentioned. 

\subsection{Application to settlement data}

Using the outlined approach, we look for potential models to describe the available settlement data from the \textit{WorldPop} data set in the different regions (India, Egypt, Nigeria). We scan sets of thresholds from $l=10^{-6}$ to $l=10^2$ and optimization hyper-parameter $\alpha=10^{-3}$ to $\alpha = 10^{3}$.
The SR3 algorithm was applied with a tolerance of $10^{-2}$ and using 200 iterations (for details see \cite{Zheng2019}). 
 
From the parameter scan, we determine for which combinations of $l$ and $\alpha$ the identified models provide an $\text{AIC}<2$. 
In order to pre-select possible solutions, we only evaluate equations with this AIC at overlapping parameter pairs of $l$ and $\alpha$. 
The reason for this is, that if the dynamics in the whole region follow the same rules (or dynamical behavior), all best identified models should have the same mechanistic form. 
Therefore, in Fig.~\ref{fig:results}i, we show for which sets of ($l,\alpha$) the low-AIC regions of each ROI overlap (in yellow). 

Next, we selected unique model equations at their lowest parameter values for $l$ and $\alpha$, respectively. The parameter combinations for unique solutions of each ROIs are shown in Fig.~\ref{fig:results}i.
These unique solutions are further compared with the use of the AIC, where we also depict the error and the complexity of the found equations in Fig.~\ref{fig:results}ii. 
The optimal selected equations are highlighted by markers with red borders and the values of coefficients are shown in Fig.~\ref{fig:results}iii. 


\begin{figure}[b]
	\centering
	\includegraphics[width=8.5cm]{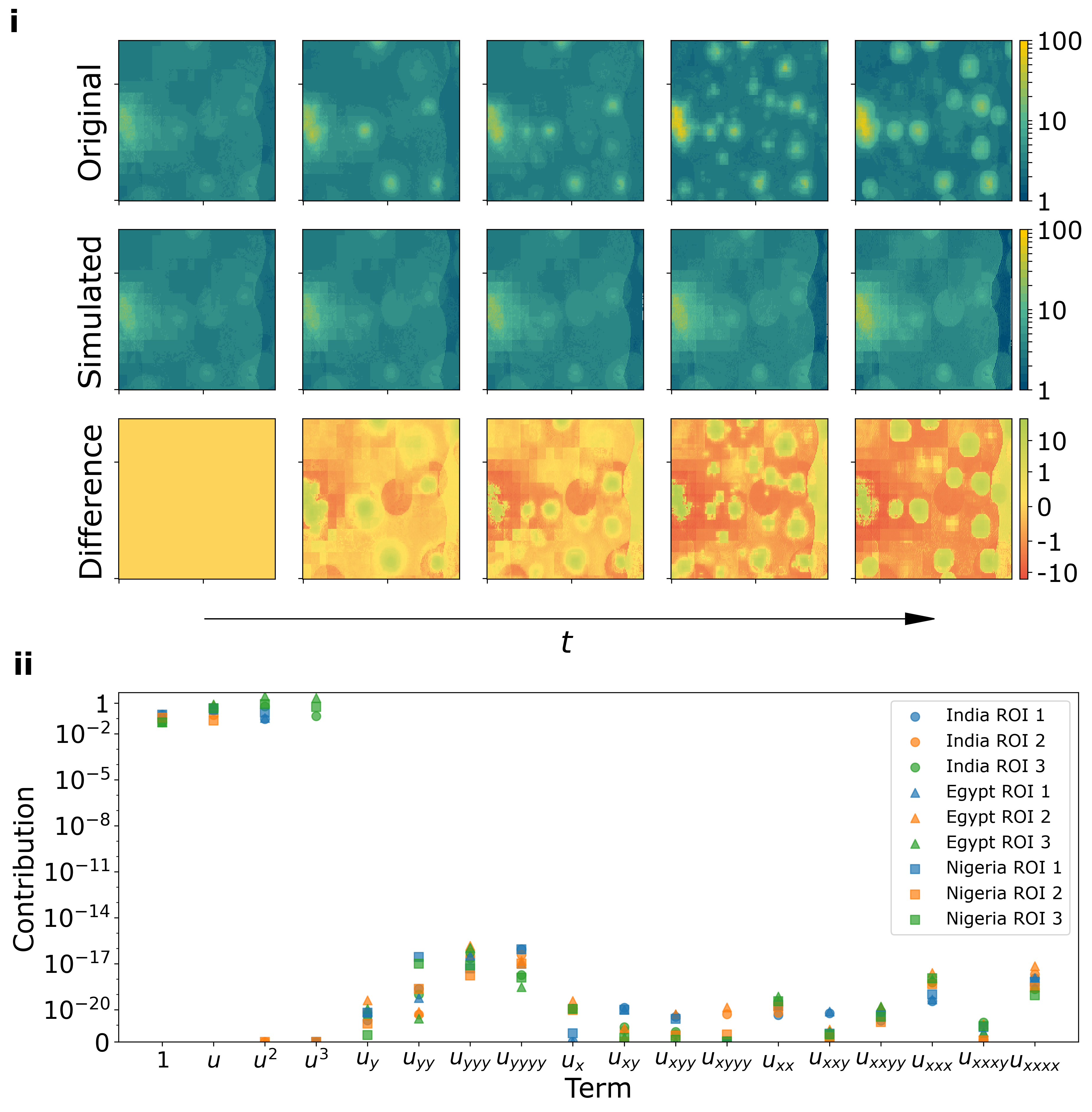}
	\caption{(i) The simulation of found models shows that the concentration in the whole region increases with time, but no patterns form. Despite being trained on the data set, the models are not able to reproduce the training data (Here ROI 2, India). (ii) Calculated contributions to the change of $u$. Even though the magnitude of coefficients is in the order of $10^{6}$ for derivatives, their contribution is negligible and lies between order of $10^{-17}$ and $10^{-21}$. }
	\label{fig:results_data}
\end{figure}

For the identified models, we analyze if they indeed reproduce the same spatiotemporal dynamics and patterns as present in the original data they were trained on. Fig.~\ref{fig:results_data}i shows a representative time evolution of the original and simulated data for ROI 2 in India. One can see that while the spatial patterns become more pronounced in the original data, this is not the case in the simulated model. This is also seen by explicitly plotting the difference between the original data and simulation. These observations are general for all analyzed ROIs (full analysis in the shared repository).

To better understand why the identified model equations do not accurately capture the original data, we look more closely at the coefficients of each model term, as well as the overall contribution of each term. We find that the assigned coefficients vary over multiple orders of magnitude, ranging from $10^{-3}$ for some production terms up to $10^{6}$ for some higher order derivatives of the diffusive terms. We calculated the actual contribution $c_j$ of each term as the value of each term $\theta_{ij}$ at a set time point $t$ (we arbitrarily chose $t=10$) and averaged over all spatial points, and then multiplied with the respective coefficient $\xi_j$,
\begin{equation}
	c_j=  \left| \xi_j \cdot \bar{\theta_j} \right| \quad \text{with} \quad
	\bar{\theta_j} = \frac{\sum_{x=0}^{X} \sum_{y=0}^{Y}\theta_{i=10\text{ }j}(x,y)}{XY}
\end{equation}
Fig.~\ref{fig:results_data}ii shows that even though the coefficients of derivative terms were multiple orders of magnitude larger than those of the production terms, their contribution is in fact negligible compared to production terms. The dominance of the terms proportional to $(1, u, u^2, u^3)$ over all terms with spatial derivatives leads to the observed overall increase in concentration while preserving the initial pattern. In conclusion, the identified models do not capture any pattern-forming mechanism as the interaction of reaction and diffusion processes is crucial (see Section~\ref{sec:settlement_pattern}).

\section{Low-data limits in model identification of pattern-forming processes}
\label{sec:sindy_AC}

One possible reason is that the SINDy method was unable to recover a reaction-diffusion model that correctly describes the settlement data due to a lack of spatiotemporal resolution and/or insufficient observation time. Indeed, it is known that model discovery with SINDy is dependent on the amount of  temporal points and the size of the time step $\Delta t$ \cite{Thaler2019, Zheng2019}. Therefore, we decided to study the limits of the SINDy method in recovering a reaction-diffusion model for low spatial and/or temporal resolution, as well as short observation times. As we found that the observed settlements followed coarsening dynamics (see section \ref{sec:settlement_pattern}), we decided to study the recovery of the Allen-Cahn (AC) equation (\ref{eq:AC_org}, \cite{Allen1979,Cross1993}) using SINDy in the low-data limit. 

\begin{align}
	\begin{split}
		\label{eq:AC}
		u_t(x,y,t)= \: &\alpha \nabla^2 u + \beta u + \gamma u^2 - u^3  \\
		\text{with} \quad &\alpha=0.1, \beta=0.5, \gamma=-0.01.
	\end{split}    
\end{align}

\begin{figure}[b]
	\centering
	\includegraphics[width=8.5cm]{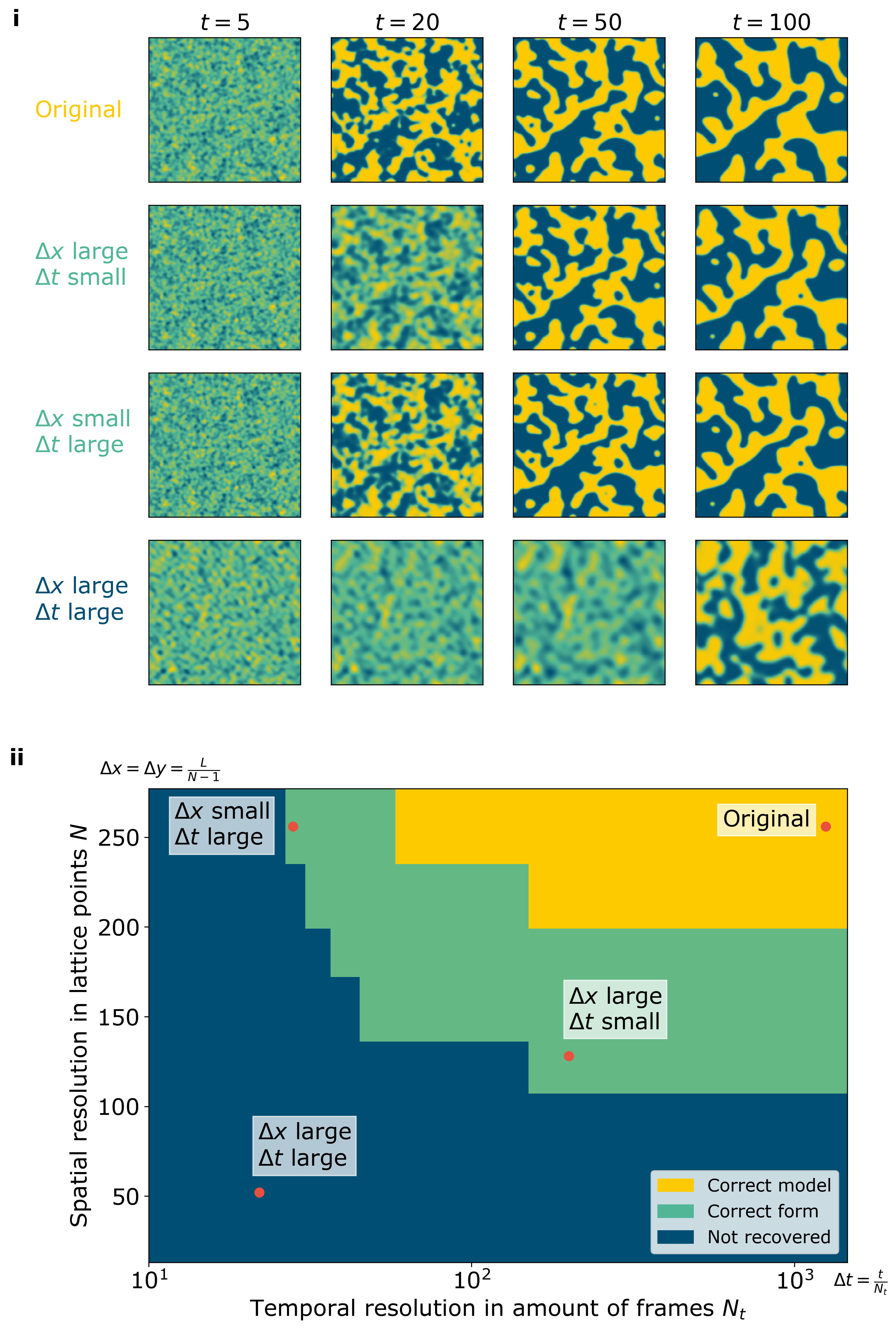}
	\caption{Recovery of the AC equation with SINDy: sensitivity to spatiotemporal resolution.
 (i) Time simulation of the AC equation, sub-sampled with different spatial and temporal resolutions, as indicated in (ii). (ii) Diagram showing for which resolution SINDy is able to recover the AC equation. Recovery is only successful for sufficiently high spatial and temporal resolution ($214 \times 214$, $N_t>80$ frames) of the simulated data set of $t=100$. }
	\label{fig:resultsAC}
\end{figure}

With this set of coefficients, the initial condition $u_{init}(x,y,t=0) \sim \mathcal{N}(0,0.01)$ (which was created once and used for all simulations), and zero-flux boundary conditions, Eq.~(\ref{eq:AC}) creates coarsening labyrinth patterns as shown in Fig.~\ref{fig:resultsAC}i (Original).

We then first sub-sampled this dataset generated by simulating the AC equation by imposing a spatial resolution of $\Delta x=\Delta y=0.39$ ($N_x=N_y=N=256$ lattice points with $\Delta x=L/(N-1)$ and $L=100$) and a temporal resolution of $\Delta t=0.08$ ($N_t=1250$ frames with $N_t=t/\Delta t$ and $t=100$). This sub-sampled dataset was then used as input to the SINDy algorithm, which was able to recover the original AC equation (\ref{eq:AC}) in the correct form with a maximum error of 3\% in the coefficients.  

\begin{figure*}
	\includegraphics[width=0.9\linewidth]{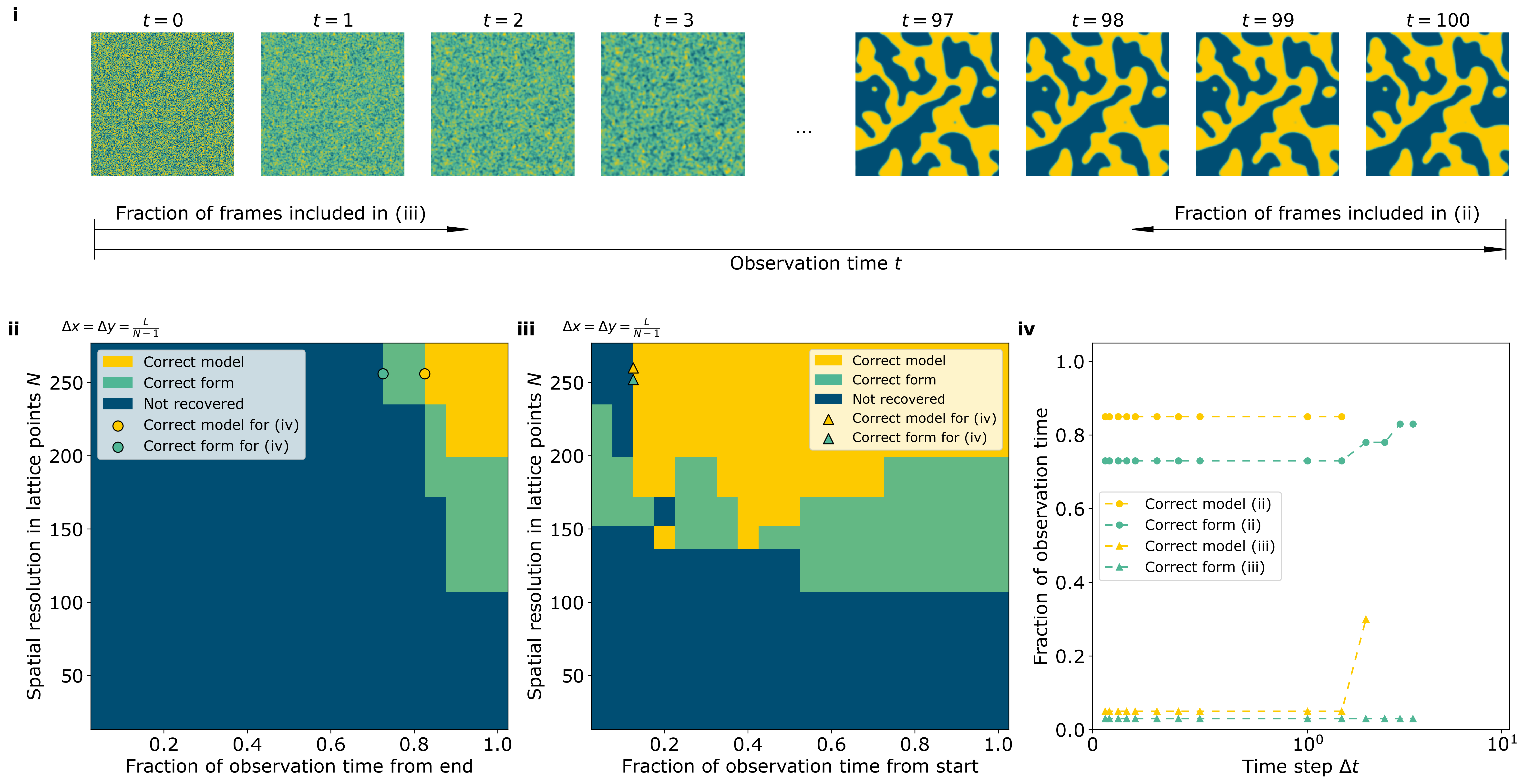}
	\caption{Recovery of the AC equation with SINDy: sensitivity to observation time.
 (i) Time series of the simulated AC equation with temporal resolution of $\Delta t=0.16$ and observation time $t=100$. (ii)-(iii) Diagrams showing for which resolution SINDy is able to recover the AC equation with changing observation time and spatial resolution. The observed time window used for model recovery is chosen at the end (ii) or start (iii) of the time series in (i). (iv) For a spatial resolution of $N= 256$ lattice points varying the temporal resolution shows that identification is less sensitive to resolution then to the amount of available data (for case (iii), the transition points are plotted with offset for visualisation).}
	\label{fig:results_time_AC}
\end{figure*}

Next, we further sub-sampled the data temporally and spatially in order to identify the limits of the SINDy approach. 
We decreased the temporal resolution from $\Delta t= 0.08$ ($N_t=1250$ frames) in 26 steps to $\Delta t= 10$ ($N_t=10$ frames). Similarly, the spatial resolution was reduced from $\Delta x = \Delta y = 0.39$ ($ N=256 $ lattice points) in 18 steps to $\Delta x = \Delta y = 7$ ($ N=14 $ lattice points).
We then compared the identified model to the original AC model, characterizing if the form of the model (correct terms) was correct, and whether it had identified the correct coefficients, see Fig.~\ref{fig:resultsAC}. This analysis shows that the SINDy algorithm is sensitive to spatial and temporal resolution. 
Only for sufficiently high spatial ($\Delta x = \Delta y < 0.46$, $N>214$ lattice points) and temporal resolution ($\Delta t < 1.25$, $N_t>80$ frames) we correctly recover the AC equation with the proper coefficients (yellow region in Fig.~\ref{fig:resultsAC}ii). For lower resolution, we are still able to recover the mechanistic form of the equation (but not the correct coefficients) in the green region in Fig.~\ref{fig:resultsAC}ii. In Fig.~\ref{fig:resultsAC}i we show how the detected mechanistic forms in this region are still able to recover the overall dynamics of the system (for low-spatial/high-temporal and high-spatial/low-temporal resolutions). However, the reduced resolution leads to coefficients smaller than those in the original AC equation, which slows down the dynamics. When decreasing the resolution even further, the optimization algorithm includes and overestimates higher order derivatives. Such models are no longer able to capture the dynamics of a system accurately (blue region in Fig.~\ref{fig:resultsAC}ii; Fig.~\ref{fig:resultsAC}i low-spatial/low-temporal resolution). 

We then wondered how the observation time affects the possibility to recover the correct model. This question is especially relevant as we only observe relatively slow changes of settlement structures over 20 years using the \textit{WorldPop} data set. We repeated the study of the recovery of the AC equation (Fig.~\ref{fig:resultsAC}) for a fixed time resolution ($\Delta t=0.16$, $N_t=625$ frames) for which we could successfully identify the AC equation (provided the spatial resolution was sufficiently high). We then redid this analysis for varying spatial resolution and observation time. The observation time was progressively reduced by only considering the first or last fraction of the original time series as input to the SINDy optimization (see sketch in Fig.~\ref{fig:results_time_AC}i). Note that in this way we are also reducing the total amount of input data.
The results of this analysis are shown in Fig.~\ref{fig:results_time_AC}ii-iii.

As expected, the AC equation can no longer be recovered when reducing the observation time and/or spatial resolution below a certain threshold. Interestingly, we find that the system is very sensitive to observation time when using the later stages of the dynamical evolution (Fig.~\ref{fig:results_time_AC}ii), while this is much less the case when using the initial part of the dynamical evolution of the AC equation (Fig.~\ref{fig:results_time_AC}iii). This illustrates that the observation time required to correctly recover the underlying model equation strongly depends on when one measures the system dynamics. In particular, our analysis shows that it is best to capture as much as possible of the dynamical changes at the relevant time scales. In this case, much of the initial patterns form quickly at the start, while later the patterns coarsen only slowly. 

Finally, we then also investigated how sensitive these findings related to observation time were with respect to time resolution. We fixed the spatial resolution at $\Delta x=\Delta y=0.39$ or $N=256$ lattice points and determined the critical thresholds in terms of observation time for correct model identification for varying time resolution $\Delta t$. Fig.~\ref{fig:results_time_AC}iv shows that successful model recovery is less sensitive to time resolution (can be varied over 2 orders of magnitude) than to observation time (both duration and exact timing). 

This also shows that there is a limit to the additional information that can be provided by higher spatial and temporal resolution if the duration and/or timing of the observed time window is not well chosen for successful model recovery. In the case of the settlement data set under study here, this analysis suggests that it is plausible that the observed changes in population density in the provided data set are inadequate for proper model identification due to the too short observation time compared to the relevant time scales over which settlements develop. 

\section{Discussion and Conclusion}
\label{sec:discussion}

The goal of this work was to not only theoretically describe the possible role of simple pattern-forming mechanisms in the development of urban structures (in our case settlements) of the Global South, as has been done before by \cite{Pelz2019,Friesen2019}, but to provide an unbiased approach to identify such models directly from data.

In order to do this, we selected three representative regions of the Global South from India, Egypt and Nigeria and analyzed the occurrence of regularity in settlement structures in these. 
Using this data, we selected smaller regions of interest (ROIs) in the spatiotemporal data set \textit{WorldPop} \cite{Worldpop} of population density distributions. 

Following this, we extended the ideas of \cite{Pelz2019}, motivating and providing a new theoretical point of view on pattern-forming mechanisms in rural, agriculturally dominated settlement structures.
We argued that together with features of regularity (as suggested by \cite{Henn2020,Tarnita2017,Pringle2017}), pattern-forming mechanisms could be responsible for the emergence of settlement structures. 
This extension can be a starting point in critically evaluating urban modeling approaches that strive for more complexity over generalization.
Here, we substantiated this claim as we observed the suggested behavior in actual population density patterns, while also showing that the characteristic length of patterns resulting from settlements follows a power law, similarly to coarsening patterns in e.g.\ Allen-Cahn or Cahn-Hilliard models. 

We then introduced the SINDy \cite{Brunton2016,Rudy2017} method together with the AIC \cite{Mangan2017,Akaike1973}, allowing us to derive and investigate spatiotemporal models for the dynamics of population density patterns. However, using the SINDy method, we were not able to identify simple pattern-forming mechanisms directly from selected ROIs of regions in the Global South. The found equations were neither sparse nor represented known pattern-forming mechanisms from literature. 
The assigned coefficients differed in multiple orders of magnitude between production terms ($10^{-3}$) and diffusion terms ($10^{6}$). 

As a result, it seems necessary to change the target of optimization from solely evaluating the coefficients to targeting the actual contribution of terms, which e.g. has been recently suggested by \cite{Naozuka2022}.
Additionally, the configuration of our SINDy approach does not include any time or space dependency of parameters, as suggested by \cite{Rudy2019}, which can prevent us from capturing important dynamical behavior of settlement systems in the Global South.
The found models show that the models are not able to recreate training data.
An analysis of term contributions has revealed that the model dynamics are dominated by the production terms and cannot be understood as pattern-forming mechanisms.

Following this unsuccessful application, we identified and studied challenges of model identification in spatiotemporal data sets considering the quality and availability of data.
We suggested that the used data set had too low spatial and temporal resolution or an insufficient observation time
Subsequently we studied this question with a sensitivity study of SINDy towards low-data limits. 
We show that SINDy (here PDE-FIND) is sensitive to spatial and temporal resolution while identifying that observation time and the observed dynamics have a significant influence on the recovery of underlying dynamics.
We see that when fast dynamics of pattern formation are captured lower spatial and temporal resolution and a shorter observation time are required to correctly identify the model.
When only observing slow dynamics (as seen in our settlement data) model identification is more sensitive towards limited spatial resolution and requires longer observations times in order to correctly identify the AC equation.  
Hence, we need to closely follow the rapid improvements in data acquisition with satellite imagery, which would provide us with sufficiently good data in resolution, while simultaneously and more importantly provide us with longer observation times.
If such challenges are overcome this work provides a ready-to-use framework to discover pattern-forming mechanisms in settlement development.

Moreover, the structure of the data should be adjusted. Currently the \textit{WorldPop} data set does not allow for uninhabited areas with a population density of 0. 
Here, \textit{WorldPop} itself is developing an improved data set, where population densities and built-up areas are mapped. 
At the moment of publication this data set only contains a single time point but, when extended, it will provide new opportunities to study our question.
Furthermore, the available data sets limit us to only a single observed variable introducing a strong assumption when considering models. 
Here, the application of methods coming from Koopman theory, e.g. delay embedding \cite{Bakarji2022,Champion2020}, could pose an interesting line of work that can provide 'hidden' variables, adding information for the optimization and resulting in model of higher dimensions (as has been recently attempted for synthetic data from a shallow-water model \cite{Ouala2023} or spatiotemporal Lotka-Volterra model \cite{Lu2022}).  

In conclusion, we have provided an initial framework for the evaluation and identification of the role of simple pattern-forming mechanisms in the development of settlement structures.
So far, the efforts were unfruitful to provide model equations describing such mechanisms.
However, we developed a possible theoretical motivation and were able to identify the major challenges of model identification in low-data limits in pattern-formation.

\section*{Data and Code Availability}
All calculations, simulations and graphs are done in \textit{Python}.
For SINDy, we use the package \textit{PySINDy} \cite{deSilva2020,Kaptanoglu2022} and for simulations, we developed a simple forward-Euler solver. 
All algorithms are available in our \textit{Gitlab} repository (\url{https://gitlab.kuleuven.be/gelenslab/publications/settlements.git}).
Furthermore, raw data and algorithms are archived via \textit{RDR} by KU Leuven under the link XXX.

\begin{acknowledgments}
The work of the author J.F. is funded by the LOEWE Program of Hesse State Ministry for Higher Education, Research and the Arts within the project "Uniform detection and modeling of slums to determine infrastructure needs".
We also want to thank Nikita Frolov for his input and constructive discussions. 
\end{acknowledgments}


\appendix

\section{Formulation of reaction-diffusion equations}
\label{app:rd_equations}

With the definitions from Section~\ref{sec:settlement_pattern} we formulate balance equations for the respective agents $u',v'$: 
\begin{align}
\begin{split}
    \dot{N}_u' &= \frac{\partial}{\partial t} \int_{A} u' \,dA = \int_{A} \hat{U} R\: f(u',v') - \oint_C \bm{J}_u' \cdot \bm{n} \,dC , \\
    \dot{N}_v' &= \frac{\partial}{\partial t} \int_{A} v' \,dA = \int_{A} \hat{V} R\: g(u',v') - \oint_C \bm{J}_v' \cdot \bm{n} \,dC. 
\end{split}
\end{align}
Here $N$ describes the amount of population or agriculturally used area in the finite area $A$ and accordingly $\dot{N}$ describes the change in the whole area, whereas $u'$ and $v'$ describe local changes. 
The long-distance effects are a product of the reaction terms $f(u,v)$ or $g(u,v)$ and the reaction rates $\hat{U}R$, $\hat{V}R$. 
Here, $u:= u'/\hat{U}$ and $v:= v'/\hat{V}$ are dimensionless by division with reference or maximum densities $\hat{U},\hat{V}$.

Similarly to \cite{Pelz2019}, the short-distance effects are also driven by a density gradient which can be modeled with Fick's first law. 
By applying Gauss' theorem, we get the two reaction diffusion equations: 
\begin{align}
\begin{split}
    u_{t} &= \hat{U}R\: f(u,v) + D_u \nabla^2 u, \\
    v_{t} &= \hat{V}R\: g(u,v) + D_v \nabla^2 v. 
\end{split}
\end{align}
With the additional dimensionless transformations $t:=Rt'$, $\bm{x}=\bm{x'}\sqrt{R/D_u}$,$D:=D_v/D_u$ we derive the dimensionless standard form of reaction-diffusion equations: 
\begin{align}
\begin{split}
    u_t &=  \nabla^2 u + R f(u,v)\\
    v_t &= D \nabla^2 v + R g(u,v)
\end{split}
\end{align}

\section{Linear stability analysis}
\label{app:LinStabRD}

As done in \cite{Pelz2019} and following  \cite{Cross1993} we perform a linear stability analysis around the linearized state of Eq.~(\ref{eq:RD_general}) with $u=U+\delta u \, , u=V+\delta v$ with the homogeneous solutions $U,V$, 
With the the perturbation ansatz $\delta u = \mathcal{R}[\delta \hat{u}\,\text{exp}(\sigma t + i\bm{k}\bm{x})]$ or vice-versa with $v$, we derive an eigenvalue problem with the eigenvalue $\sigma$, the Kronecker delta $\bm{\delta}$, $\bm{u}=(u,v)$, the Jacobi $\bm{J}(f,g)$ and $\bm{D}=0$:

\begin{align}
\label{eq:eigenvalue}
    \left(\sigma \bm{\delta} - \bm{J}(f,g)\right)\delta \bm{\hat{u}} &= 0, \\ 
    \rightarrow \quad \sigma^2 - \bm{J}(f,g)\bm{I} \, \sigma + \text{det}\left(\bm{J}(f,g)\right) &= 0.    
\end{align}

Solving the eigenvalue problem results in two conditions for the Jacobi matrix $\bm{J}(f,g)$, 
\begin{align}
    \bm{J} = \begin{pmatrix}  f_u\big|_0 & f_v\big|_0\\ g_u\big|_0 & g_v\big|_0  \end{pmatrix}.
\end{align}
that lead to instability, 
\begin{align}
    f_u\big|_0 + g_v\big|_0 &< 0 , \\
    \text{det}\left(\bm{J}(f,g)\right)=g_v\big|_0\,f_u\big|_0  -  g_u\big|_0\, f_v\big|_0 &>0 .
\end{align}
As described in \cite{Pelz2019} the only reasonable formulation of the Jacobi matrix is: 
\begin{equation}
    \bm{J} = \begin{pmatrix}  + & + \\ - & -  \end{pmatrix}.
\end{equation}
Other forms where the column-wise signs are the same results in concentrations spatially in phase and the form with row-wise same signs only the shown can be suitably used as shown in Section~\ref{sec:settlement_pattern}.

As Turing patterns can arise due to diffusion, we as well study the short-distance effects. 
We can reformulate diffusion as a product of specific energy $k_B T$, with the Boltzmann constant $k_B$ and temperature $T$, and the mobility $\mu$. 
At constant $T$, the ratio $D=\mu_v /\mu_v$ results in, 
\begin{align}
    \bm{B}:=\bm{J}(f,g)-\bm{D}k \quad \text{with} \quad \bm{D}=\begin{pmatrix}  1 & 0 \\ 0 & D  \end{pmatrix}
\end{align}
and allows to rewrite the eigenvalue problem in Eq.~(\ref{eq:eigenvalue}) to: 
\begin{align}
 \left(\sigma \bm{\delta} - \bm{B}(f,g)\right)\delta \bm{\hat{u}}=0.
\end{align}
This results again in two conditions for the Jacobi: 
\begin{align} 
    (f_u\big|_0 + g_v\big|_0 - k^2(1+D) &< 0 , \\ \label{eq:condition2_B}
    \text{det}(\bm{B})=(f_u\big|_0 - k^2) (g_v\big|_0-k^2)  -  g_u\big|_0\, f_v\big|_0 &> 0 .
\end{align}
Turing instability is achieved when the condition Eq.~(\ref{eq:condition2_B}) is violated, resulting in the necessary condition for the diffusion induced instability: 
\begin{equation}
    D\,f_u\big|_0  + g_v\big|_0 >0 \quad \rightarrow \quad f_u\big|_0 > -\frac{g_v\big|_0}{D}
\end{equation}

\section{Region Information}
\label{app:DataLoc}
Here we attach the geographical data of the regions used to demonstrate the workflow of our method, see Table~\ref{tab:coordinates}. The coordinates are given in decimal degrees in reference system WGS84.
\begin{table}[h]
	\caption{Coordinates of the regions of interest. \label{tab:coordinates}}
	\begin{ruledtabular}
		\begin{tabular}{llcccc}
			Region&	& West & South &East&North\\
			\colrule
			India   &		    & 75.3855	& 28.8265 & 77.4804  & 30.6380 \\
			& ROI 1     & 76.0086   & 29.4155 & 75.8528  & 29.5319 \\
			& ROI 2     & 76.7096   & 29.7066 & 76.5538  & 29.8230 \\
			& ROI 3     & 77.2547   & 30.2305 & 77.0990  & 30.3469 \\
			Egypt   &		    & 29.8650	& 29.9671 & 32.1010  & 31.8803 \\
			& ROI 1     & 30.3377   & 30.7896 & 30.1804  & 30.9044 \\
			& ROI 2     & 31.1260   & 30.6174 & 30.9686  & 30.7322 \\
			& ROI 3     & 31.5200   & 31.2488 & 31.3627  & 31.3636 \\
			Nigeria &           & 7.3774    & 11.1881 & 9.3336   & 13.1056 \\
			& ROI 1     & 8.2020    & 11.9172 & 8.0646   & 12.0492 \\
			& ROI 2     & 8.2708    & 12.5114 & 8.1333   & 12.6434 \\
			& ROI 3     & 8.9580    & 12.1152 & 8.8205   & 12.2473 \\
		\end{tabular}
	\end{ruledtabular}
\end{table}

\section{Calculation of ANN with moving windows of different sizes}
\label{app:MW}

\begin{figure}
	\centering
	\includegraphics[width=8.5cm]{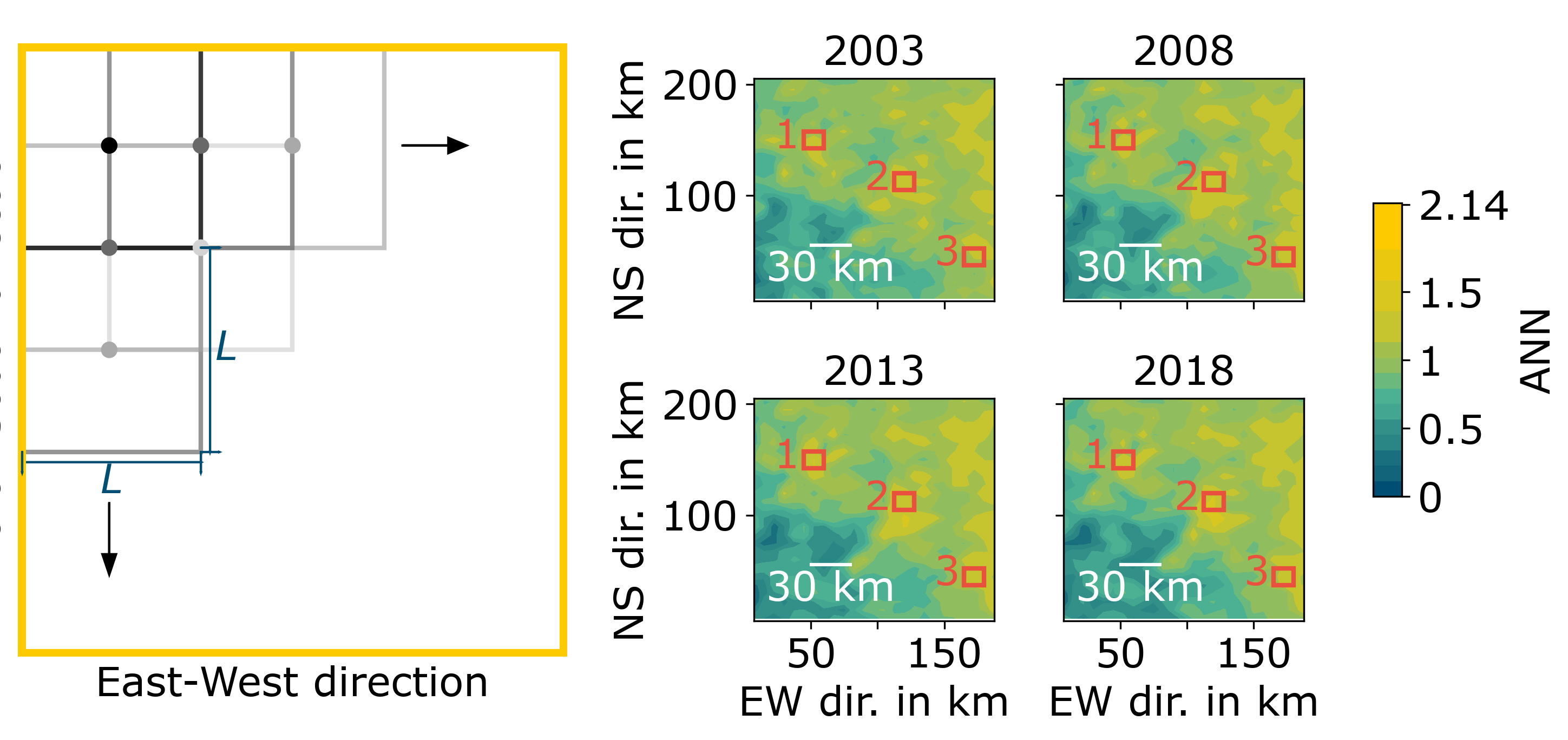}
	\caption{Moving window method for ANN calculation with changing $L$ the window side length and a part of the results for the Punjab region in India from year 2003. Similar figures for all regions, years and 10 window sizes between $L=5$km and $L=50$km can be found in the repository.}
	\label{fig:ANN_method}
\end{figure}

In order to select suitable excerpt sizes from our selected \textit{WorldPop} data sets from Tab.~\ref{tab:coordinates}, we follow a similar approach as in \cite{Henn2020}.
With varying window sizes, starting from square windows with side lengths $L$ of 5 km up to 50 km, we scan over the respective data set calculating the ANN while moving the windows in North-South or East-West direction by $L/2$ (see Fig.~\ref{fig:ANN_method}).
We calculate the ANN as in \cite{Clark1954},
\begin{equation}
    \text{ANN} = \frac{\frac{\sum_{i=1}^{N} d_i}{N}}{\sqrt{\frac{S}{N}}}
\end{equation}
with $d_i$ the distance of a settlement to the next nearest settlement, $N$ the total amount of settlements in a window area $S =L^2$.
The ANN evaluates the regularity of a point pattern and assigns a value between 0 and 2.14 describing the distribution of settlements (0– clustered, 1– random, 2.14– regular).
With this we generated contour diagrams of ANN in the selected regions over time allowing us to estimate a 'characteristic' window size where regularity is dominating. 
By visual inspection we first selected the suitable window size to be 15 km and three respective ROIs of this size which have the most regular distribution over the observation time from our contour diagrams (see Fig.~\ref{fig:ANN_method} for a part of the obtained results). 

\section{Calculation of the characteristic length}
\label{app:cl_calc}

We calculated the characteristic length following \cite{Puri2004} and \cite{Konig2021} which define it as follows,
\begin{equation}
	\ell(t) = \frac{2\pi}{\int qp(q,t)dq}.
\end{equation}

Here, $q$ describes the modes or waves lengths of a Fourier analysis of a spatial system with their respective probability of occurrence $p(q,t)$, evaluated in all directions or over $2\pi$ at every point. 
To simplify the analysis, we only scan four directions: horizontal, vertical and two diagonal directions for every point.
\bibliography{bibliography.bib}
\end{document}